\documentclass{article}
\usepackage{arxiv}

\usepackage[utf8]{inputenc} 
\usepackage[T1]{fontenc}    
\usepackage{hyperref}       
\usepackage{url}            
\usepackage{booktabs}       
\usepackage{amsfonts}       
\usepackage{nicefrac}       
\usepackage{microtype}      
\usepackage{lipsum}             
\usepackage{graphicx}

\title{Smartphone camera based pointer}

\author{ Predrag Lazi\'{c}\\ 
        Research Computing Support Services\\
	Division of Information Technology\\
	University of Missouri\\
	Columbia, MO 65211, USA\\ 
        \texttt{plazicx@gmail.com} \\
}

\renewcommand{\headeright}{P. Lazi\'{c}}
\renewcommand{\undertitle}{A PREPRINT}

\begin{document}
\maketitle

\begin{abstract}
Large screen displays are omnipresent today as a part of infrastructure for presentations and entertainment. Also powerful smartphones with integrated camera(s) 
are ubiquitous \cite{aiming_distance}. However, there are not many ways in which smartphones and screens can interact besides casting the video from a smartphone. In this paper, we 
present a novel idea that turns a smartphone into a direct virtual pointer on the screen using the phone's camera. The idea and its implementation are simple, robust, efficient and fun to use.
Besides the mathematical concepts of the idea we accompany the paper with a small javascript project \cite{mobiletvgames} which demonstrates the possibility of the new interaction technique presented as a massive multiplayer game in the HTML5 framework.
\keywords{ large display \and pointer \and interaction techniques \and smartphone camera \and augmented reality \and HTML5 \and Phaser3 \and computer game }
\end{abstract}

\section{Introduction}

\label{intro}

With the LED technology \cite{nobel_blue_LED} high quality projectors and large screens  are available at low prices today. Also, screens are rather big 
with astonishing image quality -- 4k resolution, being the standard today, at which a human eye can not distinguish individual pixels from a normal viewing distance. Smartphones are being constantly improved but it is not likely that their screen dimensions will increase beyond pocket size. On the other hand cameras in smartphones, along with the CPUs, GPUs and even NPUs are already incredibly powerful and will only get better with time.
It would be interesting if users could use their smartphones in a simple and intuitive way to interact with the available private or public large screens. Some solutions in this direction have been tried out but most of them remained on a purely academic level -- mostly as concepts or proofs of principle \cite{registered_screen_proof}. What we present in this paper is a simple solution that enables seamless interaction with the screen using only the smartphone with a camera and a (network) connection between the two. We describe the motivation for developing this technique, current state of the art, and the technical implementation of our solution. We discuss briefly the details of the implementation and give guidelines for further development of the technique especially along the lines of machine learning and neural networks. The demo product of our implementation is the computer game done in JavaScript and HTML5 within the Phaser3 framework \cite{website_phaser}. The game is an homage to the legendary NES Duck Hunt game \cite{website_DuckHunt} with distinction that our game is a multiplayer one intended for tens or even hundreds of players. \\
We believe that our solution has a potential to spark the revolution in finding many new, creative and fun ways of multiple users interacting with large screens.

\subsection{The Motivation} 
The idea for the project of camera based pointer emerged from the frustration while giving presentations at conferences. Namely, using the pointer during presentation is of utmost importance to guide the eyes of the audience and to communicate ideas efficiently. However, of the three existing options for pointer each one has its drawbacks. \\
\begin{itemize}
\item \textbf{Laser} pointer (red or green) has an advantage that it works as a real pointing device following the natural, innate, human sense for aiming. It works from great distance and is clearly visible when used on a projector screen. However, laser pointer, in our opinion, due to its large brightness creates a new separate layer on top of the presentation slide and is not immersed in the presentation. Also, it can not be used to click on things in presentation. Using the laser on a TV screen is usually impossible due to a reflectivity of the TV screen while specular reflection off a screen surface also presents a health hazard.  \\
\item	Another device used as pointer in presentation is the so called \textbf{air mouse} -- the device that connects to the computer running the presentation via dongle (typically) and is using accelerometers to move the real mouse pointer within the presentation. In this way the pointer is seamlessly immersed in presentation, can be used for clicking and even for drawing. However, the large drawback of the air mouse, and the main cause of frustration, is the fact that it generates relative movement of the mouse pointer (exactly as in the case of an ordinary computer mouse). Gathering information from the built in accelerometers moves the mouse pointer in given direction but the alignment of the mouse pointer and user's hand holding the device is not guaranteed. It often gets misaligned and following the natural tendency of having the laser-like aiming pointer user has to align the two by pushing the pointer to the edge of the screen. Misalignment is particularly frustrating when user tries to draw (underline, cross out or circle) something in the presentation slide. This loss of natural aiming interaction is the largest drawback of the air mouse type pointers. Also accelerometers are presently not sufficiently accurate. \\
\item	Finally, one can always choose a \textbf{simple stick} to serve as a pointer which besides limiting the users freedom of movement also casts a shadow and does not enable interactive clicking on the presentation.\\
\end{itemize}

With all the advantages and disadvantages of the available pointer solutions one can immediately conclude that the ideal pointer is the one that has a laser like aiming property but is at the same time seamlessly immersed in the presentation.\\

We decided to develop such a pointer solution using a readily available smartphone with camera. The requirements of the pointer screen interaction are nicely described in \cite{Boring_2009} \textit{The interaction techniques need to be efficient, enjoyable and easy to learn}. For that reason, besides making another proof of principle concept we have developed a fully functional game since the game requires such interaction properties to be playable. Moreover, since the pointer--screen interaction already requires a client--server structure we also made the game a massive multiplayer one.\\
In general, in the paper we reference computer games a lot since they present the ultimate multimedia experience.

\subsection{State of the Art}
The state of the art can be distilled in the sentence following the one on requirement for the interaction technique, stated above, from \cite{Boring_2009} where the authors conclude \textit{Unfortunately, each of them has its limitations in terms of accuracy, operation speed, efficiency and ergonomics.} We conclude that the situation has not changed since then  -- from studying the literature and simply by observing that interaction between the smartphones and large screens does not exist around us. Most of the interaction with screens takes place through the remote controllers. Here we give a brief overview of the several directions and attempts to develop cellphone -- large screen interaction techniques.\\
Most of the techniques were conceived in the late 2000s when the phones with cameras were becoming a standard, however the computing power of such phones was nowhere near the ones of today. It is interesting to note that only today the browsers with their implemented technologies (such as WebGL) are actually mature enough to support implementation of our simple technique fluently. \\
The attempts that are found in literature have one major drawback and it is that they are over-engineered in a sense that they in principle require too much resources to do the job for fast paced applications. Some of the solutions used the visual codes (AR markers) \cite{website_ARMarker} on the screen to detect what user is pointing at \cite{Ballagas_2005} whether as setting up an absolute coordinate system or by tagging individual objects on the screen. In the same paper the authors use cellphone camera as the standard optical mouse technique resulting in relative movement of the pointer (\textit{sweep}). \\
The solution that has some vague similarity with our idea is the so called \textit{display registration} \cite{registered_screen_1,registered_screen_proof}. In the first implementation of that approach the special color markers are drawn on the screen in order to establish the coordinate system of the camera's video from the perspective of the server (screen) which is done by calculating the homography transformation matrix \cite{hartley_zisserman_2004}. In general, that class of solutions, using visible markers, are considered sub-par, obviously for modifying the screen content with markers, and the optimal solution is sought within the framework of the so called markerless solutions.\\
 One of the markerless implementation attempts is given in Ref. \cite{Boring_2013}. That solution is based on cellphone sending images from its camera video to the server which then searches for features in them and matching them to the features of the image on the screen -- produced by server. This requires, first of all, significant communication between the smartphone (client) and the screen (server) and besides that also the high computational resources are needed resulting, among other problems, in temporal mismatch between the current content on the display and the most recent camera frame. Such solution is not suitable for example in fast shooting game, whereas the one we developed is. We reflect on a seemingly odd solutions that were developed through a historical perspective later in the paper. \\
Finally, we would like to mention the interesting attempt \cite{Jiang_2006} where the computer draws initially a pointer symbol (colored dot) somewhere on the screen while the user is holding the camera looking at the screen. Upon detecting the dot, camera sends the information about the dot's position relative to the center of the camera's view back to the screen (server running it). Server then updates the dot's position accordingly and the procedure is iterated until the two coincide. This solution is rather original but obviously has many drawbacks for practical application where the fast feedback is required while it is completely useless if the pointer dot is not wanted on the screen. Even though this idea of server continuously and iteratively adjusting the pointer position might sound counterintuitive, the motivation behind it is perfectly reasonable and it is in the same spirit of our approach to the pointer interaction. Namely, the authors wanted to create a pointer interaction technique that will not require high learning effort and they attempt to enable an intuitive pointing technique: directly pointing to the desired target.  \\
All the solutions have some of the challenges that make them unsuitable for the required screen--smartphone interaction. It should be said that however solution using a phone camera as an optical mouse was very successfully implemented in one of the first AR games (Mozzies) \cite{Lopez_2012} where user moved the cross-hair using the camera and shot mosquitoes. The game had a very natural feeling but it only used the cell phone screen.\\
It is important to understand that operating a relative movement device requires a visible pointer. For that reason, and for mouse being in a different plane than a screen (so not even remotely resembling aiming) makes a usage of a computer mouse so pleasant and natural. On the other hand direct pointing devices do not necessarily require pointer drawn for achieving interaction with the object being pointed at. This is very important in our case of implementing direct pointer where for the case of the massive multiplayer game on a single screen we simply do not draw the pointer and the game is still perfectly playable -- actually the shooting game should look exactly like that. 

\section{Camera based pointer - Implementation} 

\begin{figure}[ht]
\centering
\includegraphics[clip=true,width=0.95\textwidth]{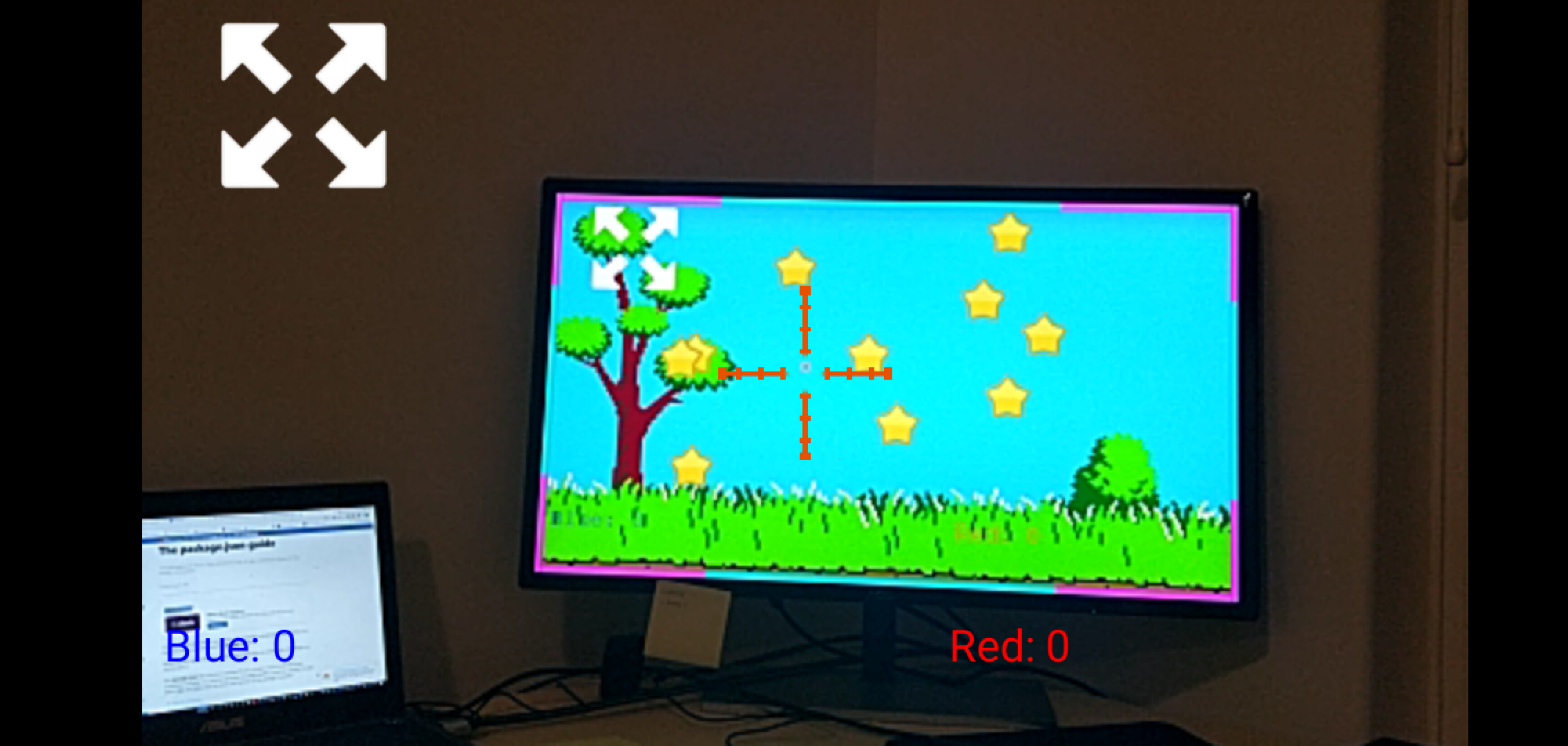}
\caption{Screenshot on the client side (smartphone) interacting with the game on the server (screen). Notice that the \textit{virtual pointer laser dot} visible almost in the exact middle of the cross-hair is actually drawn only on the server which got the information from the client about where the client is aiming at.}
\label{fig1}
\end{figure}

Current implementation of our solution might broadly be classified as a marker based solution but in discussing future work we argue that our idea can be turned into a markerless one.
Our idea of the implementation of the camera based direct pointer is surprisingly simple. In order not to obscure its simplicity we do not discuss here the most general situations when user rotates the phone besides the landscape or portrait orientations, or is standing at a very sharp angle to the screen (camera is far from being aimed perpendicularly to the screen). In the later case the rectangular screen looks as a trapezoid which is a well known effect of keystoning (appears also when projecting image) which can be compensated perfectly \cite{Sukthankar_2000}. Implementation would require some tampering with the original tracking library so we leave it for future work.\\  
In the very simple approach we expect the setup where the user (we will call it client here) sees the image on the screen (server) through the smartphone camera as shown in the screenshot of the game in Fig. \ref{fig1}.

\begin{figure}[ht]
\centering
\includegraphics[clip=true,width=0.95\textwidth]{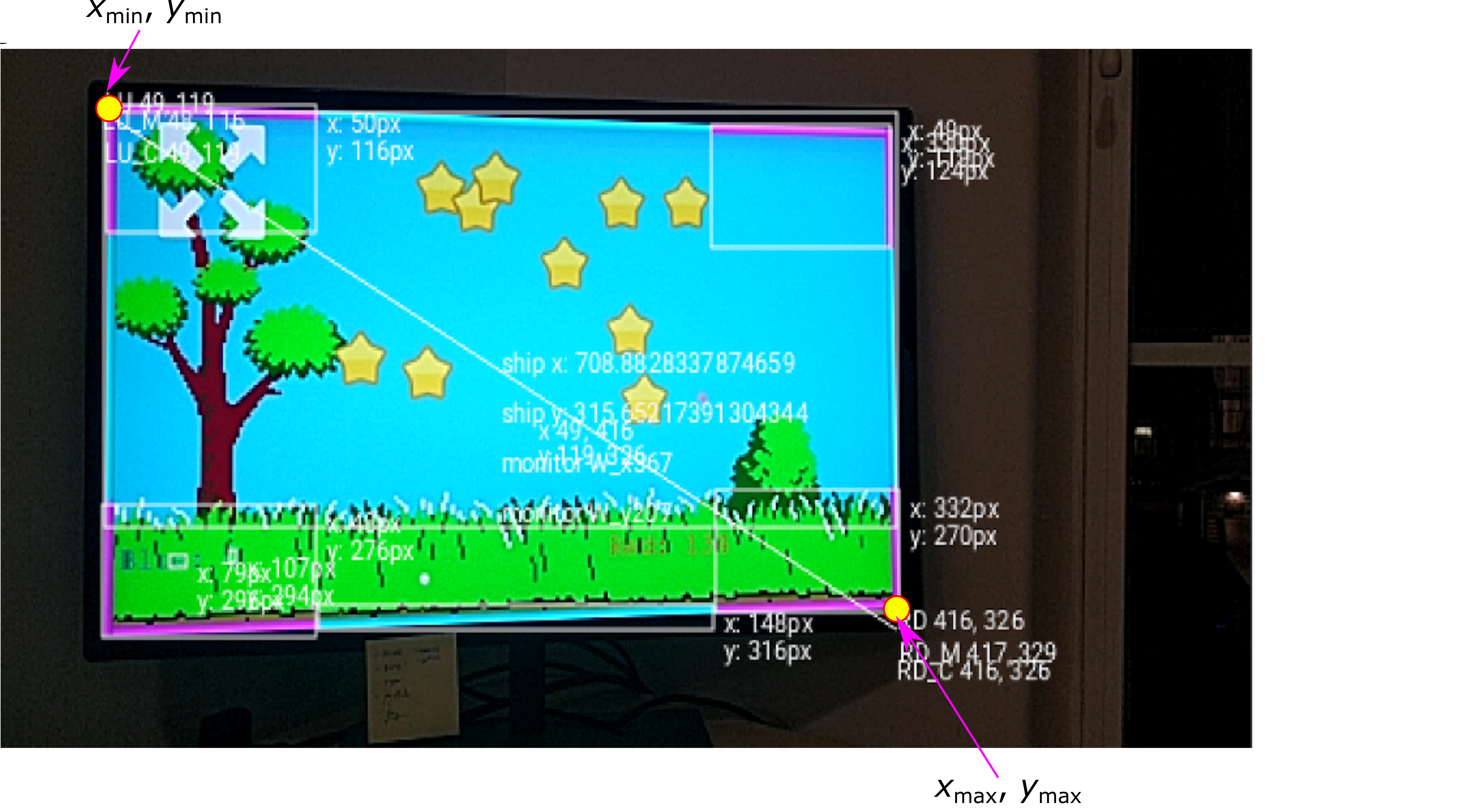}
\caption{\label{fig2} Screenshot on the client side (smartphone) interacting with the game on the server (screen) in debugging mode showing the color tracking detected features i.e. cyan and magenta colored regions (rectangles).}
\end{figure}

The most important step of the algorithm that calculates the client's aiming coordinates on the server's screen is the detection of the server's screen edges. In particular the client (camera and smartphone) has to determine two diagonally opposing corners of the server (screen). To achieve this we have used the available tracking--js library \cite{website_trackingjs} which enables several tracking modes of the features in the image (video). We chose the color tracking mode which detects features on the screen of certain color (magenta and cyan in our case). The detected features are shown in the screenshot of the debugging mode of the game in Fig. \ref{fig2}. We discuss a bit later the reasons for using two colors instead of only one and having the colors along the whole edge of the screen instead of only in the corners. Also, we will address the requirement that client sees the whole server at all times.

\begin{figure}[ht]
\centering
\includegraphics[clip=true,width=0.95\textwidth]{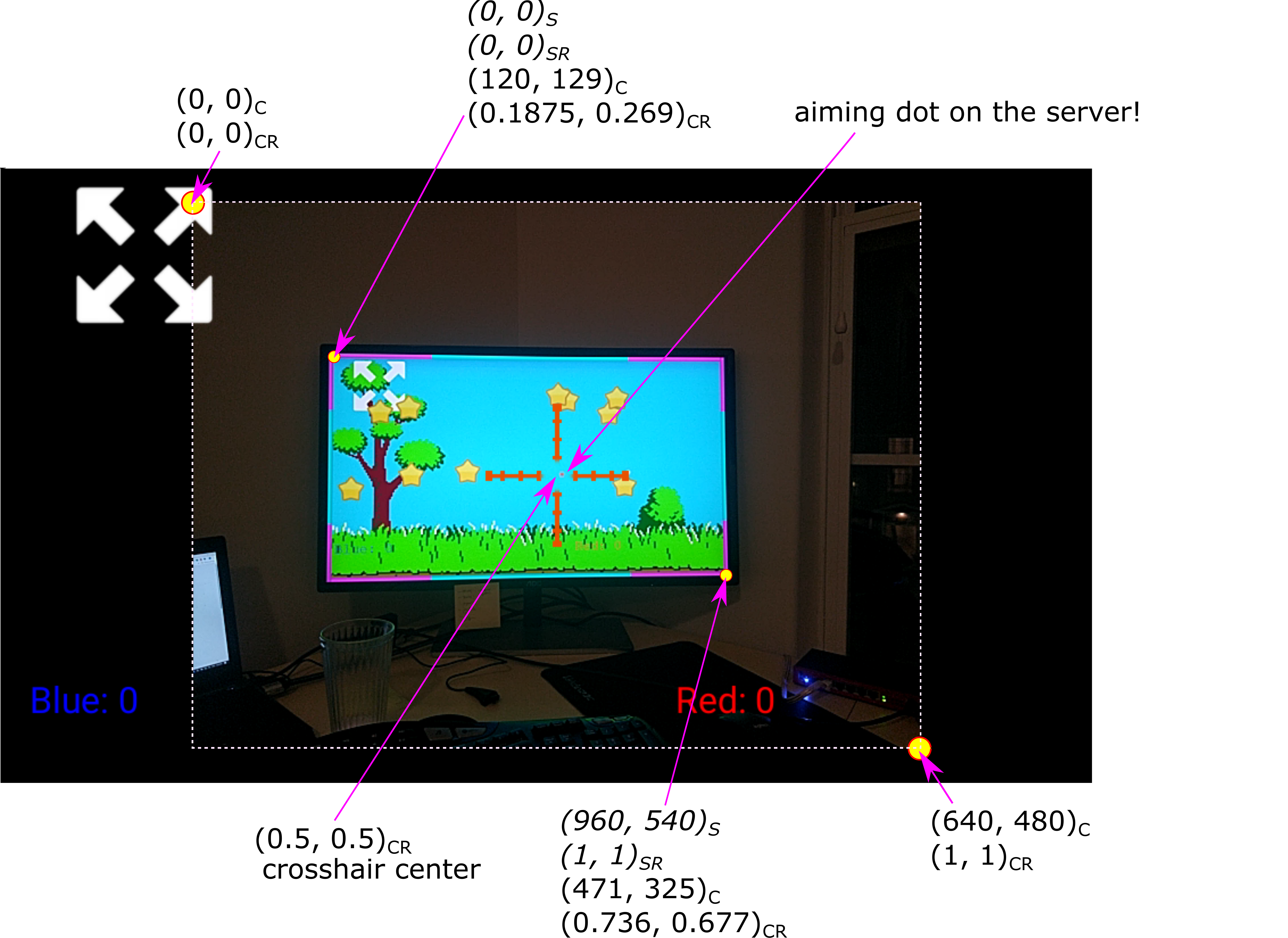}
\caption{\label{fig3} Screenshot from the client side (smartphone). The main features of the scene are marked. The crucial part for calculating aiming point in server relative and absolute coordinates is the fact that aiming point in the camera relative coordinates is $\mathrm{(0.5, 0.5)_{CR}}$.}
\end{figure}

The successful detection of the screen edges using the tracking-js library in its color tracking modes is achieved by comparing the coordinate values of the detected features. In this way  we can obtain the values of the $x_{\mathrm{min}}$, $y_{\mathrm{min}}$, $x_{\mathrm{max}}$ and $y_{\mathrm{max}}$ which determine the two diagonal corners of the screen.\\
Upon determining the boundaries of the server (screen) we can start referencing crucial points in the client (camera) view. In Fig. \ref{fig3} we show the most important features on the client's side referencing coordinates in server (S) and client (C) absolute values as well as in server (SR) and client (CR) relative values.

After establishing the coordinates of the diagonal corners of the server screen in camera relative coordinates, and knowing that the aiming point in camera relative coordinates is $\mathrm{(0.5, 0.5)_{CR}}$ we proceed to calculate the aiming point in the server relative coordinates following Fig. \ref{fig4}.

\begin{figure}[ht]
\centering
\includegraphics[clip=true,width=0.95\textwidth]{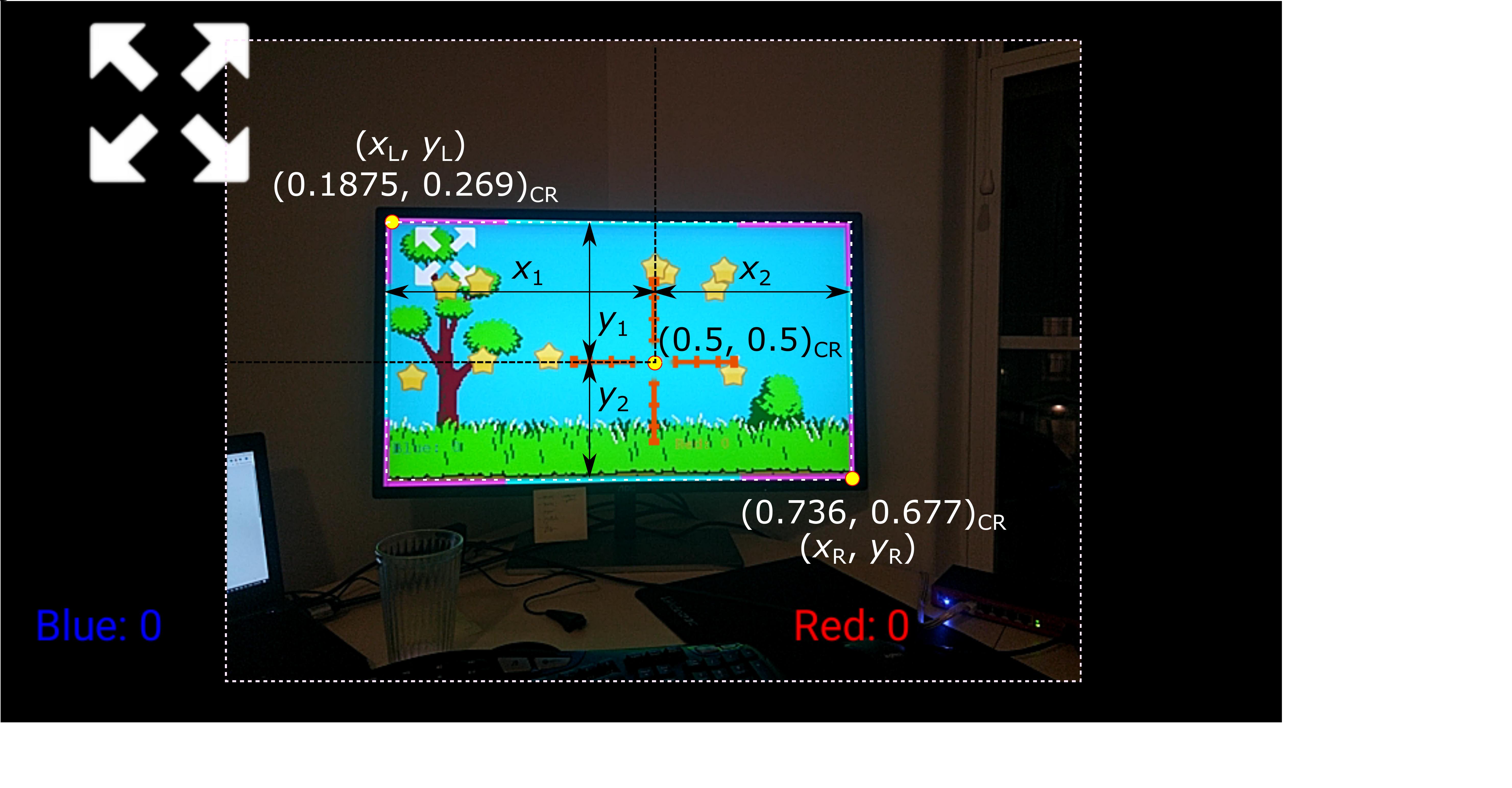}
\caption{\label{fig4} Screenshot on the client side (smartphone) marking the values $x_\mathrm{1}$, $x_\mathrm{2}$, $y_\mathrm{1}$ and $y_\mathrm{2}$ which are required to calculate the aiming point server relative coordinates.}
\end{figure}

From Fig. \ref{fig4} we arrive to the core of our algorithm which can be given as a set of equations which can be seen also as the pseudocode:

\begin{eqnarray}
\label{the_core}
 x_1=\mid 0.5-x_\mathrm{L} \mid \nonumber \\
 x_2=\mid 0.5-x_\mathrm{R} \mid \nonumber \\
 x_\mathrm{SR}=x_1 / (x_1+x_2)\\
 y_1=\mid 0.5-y_\mathrm{L} \mid \nonumber \\
 y_2=\mid 0.5-y_\mathrm{R} \mid \nonumber \\
 y_\mathrm{SR}=y_1 / (y_1+y_2)
\end{eqnarray}

In order to obtain absolute server (screen) coordinates the relative server coordinates $x_\mathrm{SR}$ and $y_\mathrm{SR}$ must be multiplied by the screen width and height resolution, respectively. Notice that it is not even required that the client knows the absolute resolution of the server's screen.\\

Even though the presented algorithm solution requires only basic coordinate transformation this simple scheme works surprisingly well! Moreover, the communication between the client and the server is very light. Namely, if the server is constantly showing the aiming point of the client(s) - i.e. if we mimic the pointer mode then, on average, the network bandwidth per client is around 2 kbps. However, if we use the implementation as in a shooting game where the user aims looking through his smartphone camera and communicates aiming coordinates only when the fire button is pressed then on average a network utilization is around 60 bps per player. This feature enables truly unprecedented ways of using this implementation where we would like to suggest one -- unseen so far. For example users going to the movie theater could play a game (a zombie apocalypse type of) where even several hundreds of clients can play the game on the same screen. Regarding the network latency, we have tested our implementation with a server in Europe (Croatia) and the game was still enjoyable to play in the USA.\\

\vspace{1cm}
\subsection{Details of the implementation}
\begin{figure}[ht]
\centering
\includegraphics[clip=true,width=0.95\textwidth]{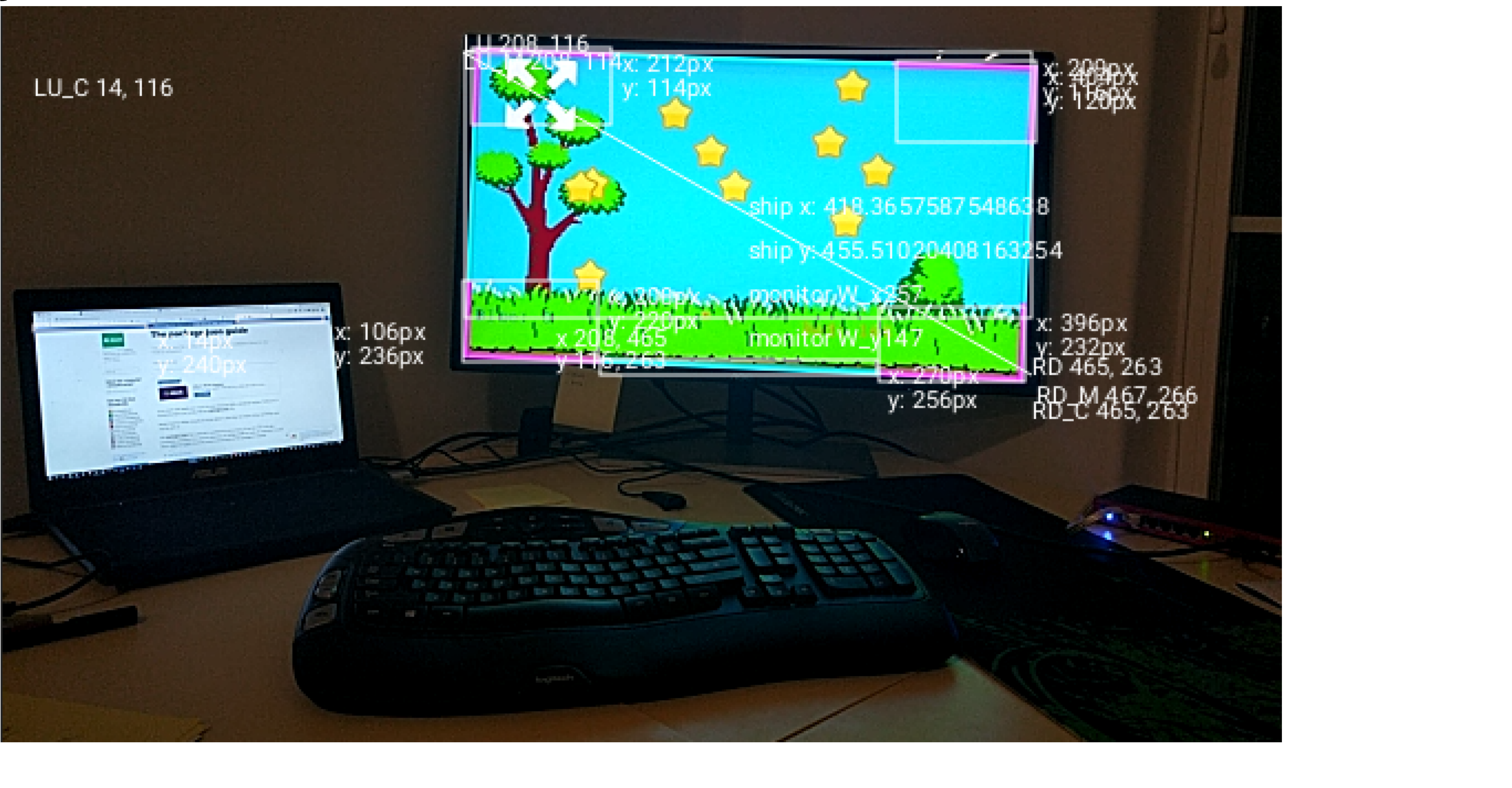}
\caption{\label{fig5} Screenshot on the client side (smartphone) in debugging mode showing tracking color detected feature. Some of the detected features are out of the scope of the server's screen -- for example the laptop's screen on the left of the server screen.}
\end{figure}

Despite the simple idea behind the algorithm its practical implementation is facing several challenges. In Fig. \ref{fig5} we show a broader image, on clients phone, in which it is visible how objects which are not part of the server's screen are being detected by the color tracking algorithm.

In the very first attempt to implement the algorithm we have simply placed two magenta squares in the two diagonal corners of the server screen. This indeed works but is very sensitive in a sense that if only one of the squares is not properly detected in a single frame the calculation of the pointer position is not possible. Our first improvement was to introduce the mono--colored edge instead of just squares in corners. This solution, even in single color was much more robust than the one with the squares in corners since even if just small parts of the edge are detected the algorithm works well. Notice that looking for the diagonal corners coordinates does not require necessarily the corners themselves to be detected (or even visible!). Detecting any part of the edge of the screen is enough to obtain \textit{extreme coordinates} - namely $x_{\mathrm{min}}$, $x_{\mathrm{max}}$, $y_{\mathrm{min}}$ and $y_{\mathrm{max}}$ from which then the corner coordinates are easily deduced. Having the color rectangles all around the edge also enabled using thinner edge since the detection of the whole shape is not necessary. In practice the needed thickness of the edge depends on the contrast and brightness of the screen, clients cameras capabilities etc. It is a variable that can be adjusted together with the shades of the colors used \cite{colors}. \\
Using a single color for the edge detection leads to the problem of tracking features of the targeted color which appear out of the scope of the server screen yielding false values of the extreme coordinates, hence prohibiting the accurate pointer coordinates calculations. To overcome this challenge we used yet another simple idea of using the two colored edge. The idea of the two colored edge is based on the assumption that the features around the server's screen will hopefully not contain both of the chosen colors. The system is now tracing two sets of the extreme coordinates one for the magenta and one for the cyan color. The values of the two sets of the extreme coordinates should match. However, in the case of detecting, for example,  a magenta feature which is out of the screen the algorithm notices that the values of the extreme coordinates for the two colors do not match. Since the cause of the value mismatch is out of the screen (non--edge feature) detection, the algorithm takes the \textit{less extreme} value of the two - meaning that if maximal $x$ coordinate of the magenta color is larger than the one of the cyan -- the cyan coordinate value will be used (i.e. the smaller one). The same is true for the minimal $x$ value - the larger minimal value (the less extreme of the two minimal values -- cyan and magenta) of the two will be used. Additionally, tracking features smaller than a certain size are simply neglected since the size of the edge rectangles is known by construction. This causes for example the blue LED of the router to be neglected (in the right lower corner in Fig. \ref{fig5}). If the system is deployed in a very colorful surroundings more than two colors can be used for the edge with the same logic of eliminating the falsely detected features coming from out of the screen. Some other features could be added to ensure stability of the algorithm such as checking for the screen ratio from the detected corners. However, the described two color edge scheme with properly selected edge thickness is rather robust. \\
Notice that features inside the screen are absolutely irrelevant since the algorithm is looking for the extreme screen coordinates so that whatever takes place inside the screen does not influence the outcome of the extreme coordinates value detection - i.e. there is no limitations in using any color in the server's screen. Using tracking of more than one color and having more than a few rectangles of the targeted color does not seem to put additional load on the CPU/GPU running the color tracking algorithm. It is interesting to mention that color tracking algorithm can be done using a WebWorker. \\
One more feature can be implemented through the usage of the rectangles of different colors for the screen's edge. Namely, the size of the rectangles can be chosen in such a way that the screen corners coordinates can be detected even if the whole screen is not visible. We can choose for example the length of the horizontal rectangles to be $1/4$ of the total width of the screen and the height of the vertical ones to be $1/3$ of the screen's height. In that case observing just one vertical and one horizontal rectangle suffices to calculate position of the screen corners. We did not implement this particular idea since in the gaming and the presentation pointer mode we expect client to see the whole screen most of the time. 

\subsection{Other challenges and different approaches}
\begin{figure}[ht]
\centering
\includegraphics[clip=true,width=0.95\textwidth]{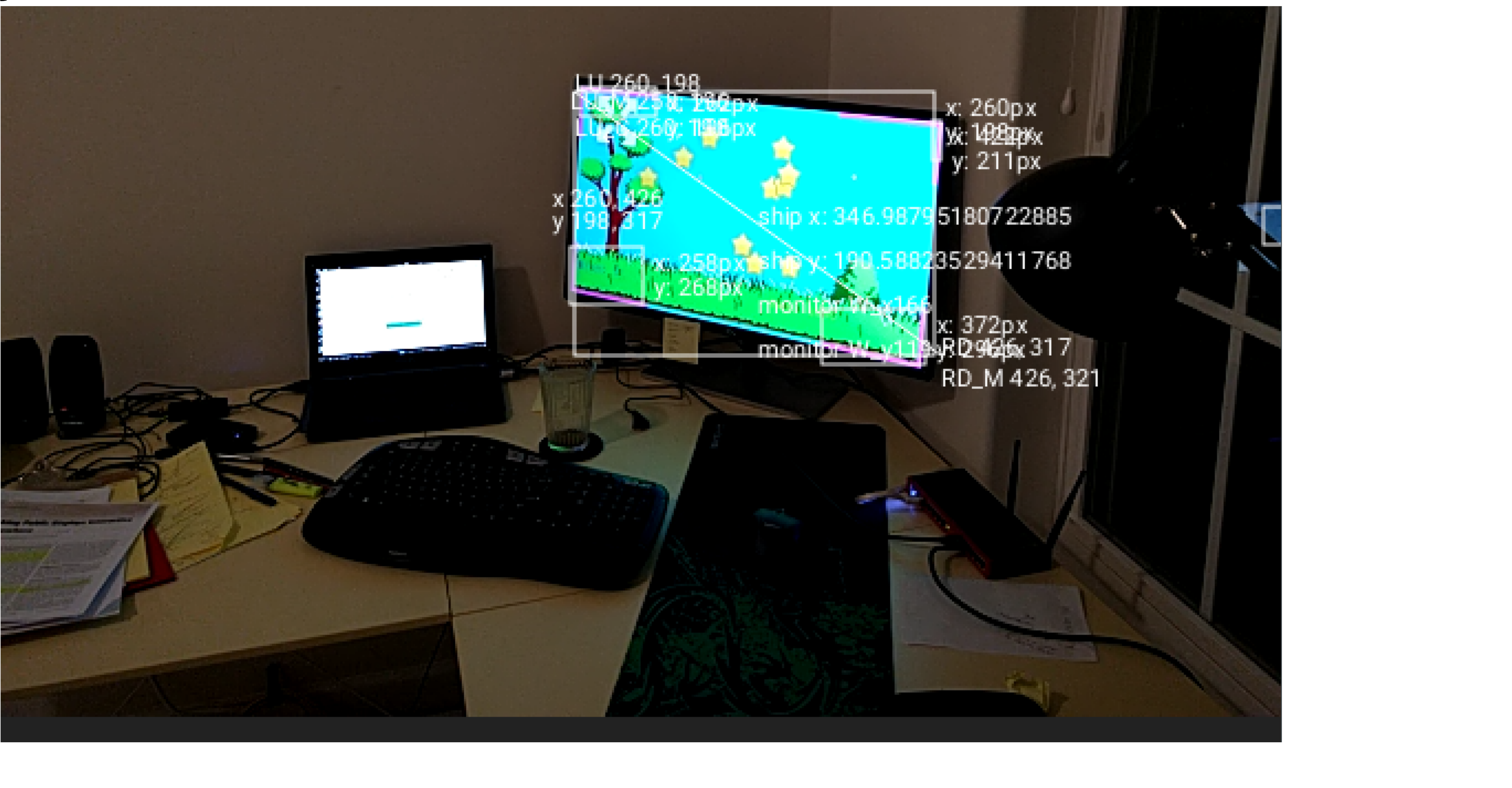}
\caption{\label{fig6} Screenshot on the client side (smartphone) clearly showing the keystoning effect.} 
\end{figure}

Besides the problems  of false detection of color features from regions that are out of the server's screen the most obvious challenge is the so called keystoning effect which can happen when the screen itself is not of the rectangular shape (typically when the projector is used) - which will not be that often as the other case when it appears due to users position relative to the screen which is very different from looking perpendicularly to the screen -- i.e. the user being at a large angle with respect to the screen. Such situation is shown in Fig. \ref{fig6} where the typical keystoning distortion is visible -- meaning that the rectangular object is viewed as trapezoid.

It is well known how the keystoning effect is compensated \cite{Sukthankar_2000} and we leave this for future work due to a need to modify a tracking library.\\
Regarding different attempts to achieve screen corners detection we have experimented with using the AR markers \cite{website_ARMarker} in the corners of the screen and even face detection with skull images placed in the corners since the tracking--js library has the face tracking implemented. Both of the attempts show as a way too demanding tasks for an average smartphone for having a very responsive gaming experience as can be achieved through the color tracking scheme.\\
Ultimately, the goal would be to achieve screen's edges detection without any sort of markers - i.e. achieving a truly markerless detection. We believe that this can be achieved using the incredible raise in the machine learning techniques in particular using neural networks for image detection. The idea of the screen detection without markers would consist of quick and efficient calibration of the pointer to detect the edges of the screen. This could be achieved by guiding the cellphone camera aiming point along the edges of the server's screen. In this process the accumulated images from the camera stream can be used to train the neural network to recognize the screen edge in the later usage of the pointer. This idea is only speculative for now but seems as the most interesting direction for future work. \\
The possibilities for improvement of the servers screen detection are numerous but we argue that the one we presented with the colored edge does not take away too much of the server's screen surface - especially when the monitor type of the screen is used. When (cheap) LED projectors are used one might need to make a slightly thicker edge. The reason for this is that typically (TV) screens have much higher contrast ratio than projectors.

\section{Evaluation}

Since the algorithm that we presented is integrated in a real HTML5 game we had no trouble testing it with users since the game interactions are the quickest one to grasp. The testers were from different backgrounds of education and of different age. What we would like to stress is that the first reaction when interacting with the system using their own smartphone through the webpage - is disbelief. Users have a feeling that they have never seen anything like this before and almost feel that something fishy is going on. The lag when using a local network is virtually nonexistent and moving the aiming dot on the screen by simply waving the smartphone is really surprisingly unexpected. To follow the analogy with the NES Duck Hunt game where the user used additional controller to shoot the ducks on the screen - the electronic gun \cite{website_ZapperNES}. It also looked very impressive back then (in the 1980s). That solution works seamlessly with CRT screens but can only detect if the user shot a target or not, i.e. it could not detect where the user is actually aiming but even such a simple solution recreated the aiming and shooting feeling so well that it was hard to believe that something like that is possible. \\ 
Continuing with the digression, after discovering the video \cite{teensduckhunt} from 2015 showing the reaction of the teenagers to playing the original DuckHunt game, published in 1984 using the Zapper gun \cite{website_ZapperNES} we expect a huge comeback of that type of games -- only that this time they will be accessible to everyone with an average smartphone. The reason why the games that require shooting with aiming are so competitive is we believe in evolutionary importance of that activity. Shooting with aiming, unlike shooting without aiming -- performed typically during celebrations in primitive societies has important evolutionary origin since it is required when hunting or defending oneself from the attack, either by animal or in war. All those activities are essential for survival, hence such a competitive edge in shooting with aiming activities. \\ 
In more technical and less subjective terms our demands for the virtual pointer were: 
\begin{itemize}
\item high accuracy
\item small latency
\item no calibration
\item massive scalability with number of users
\item no learning curve
\item fun to use
\end{itemize}
All the required characteristics were met. The only thing that was somewhat unsatisfactory was the accuracy when the user is further away from the screen \cite{aiming_distance} -- for example at a distance of 5--6 diagonal lengths of the screen. It turns out that at that distance the main challenge is the color detection if the very thin color edge is used. However, using a thicker edge makes an aiming experience from that distance a realistic life like experience -- user has to put more effort in aiming since the targets are further away. Finally the keystoning effect had almost unnoticeable influence on the accuracy of the aiming at angles of even of 60 degrees with respect to the orthogonal to the screen. Angles larger than that make no sense from the perspective of playing the game.\\
Different browsers had rather different behavior on the client side. For the server side it is almost irrelevant which browser is used since the server does not do color tracking algorithm -- i.e. server does not even need a camera. On the other hand on the client side by far the most smooth experience was in the Samsung Internet android app, followed by the Opera, Mozilla Firefox, Google Chrome and Edge.\\
Testing the game on a rather old iPhone SE using Safari (the game on iPhones works only with Safari browser since it is the only one that can get a hold of camera on the iPhone) stunned the author as it looks as the effortless task for the rather old smartphone. Different performance of different browsers is the reason for our claim that only now the browsers are relatively mature enough for this kind of applications.
Finally, instead of giving any conclusion on evaluation we invite readers to simply test the game \cite{our_github}.

\section{Historical perspective and further improvements}
\begin{figure}[ht]
\centering
\includegraphics[clip=true,width=0.95\textwidth]{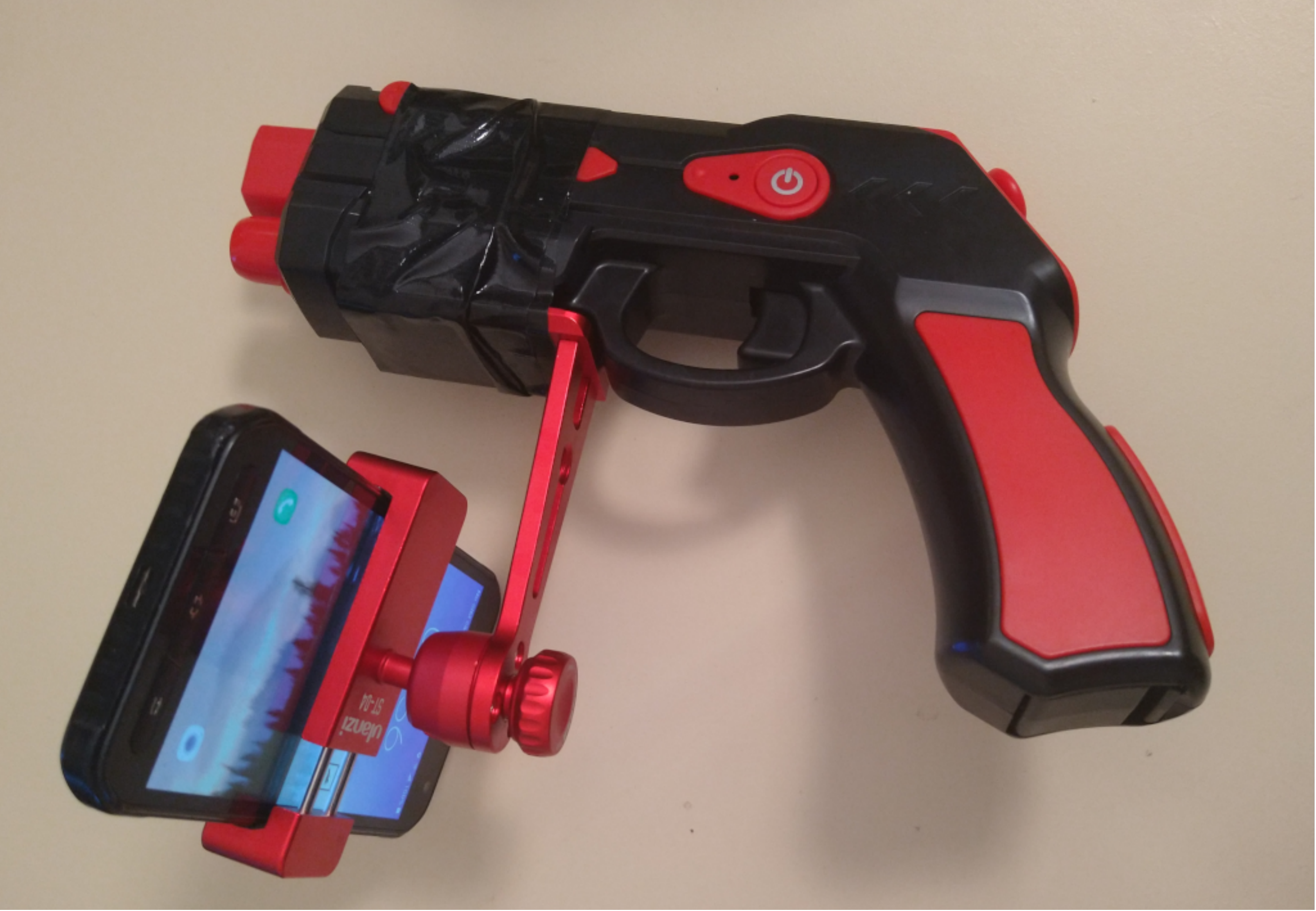}
	\caption{\label{toy_gun} Photo mount for cell phone attached to a toy gun using a duck tape. The cell phone fire button in the game is triggered via Bluetooth connection. This device requires initial calibration by adjusting the cell phone position so that aiming sight of the toy gun coincides with the cell phone aiming. Once it is achieved user can use a toy gun for a normal gun-like aiming experience and enjoy the game even more than looking through a cell phone screen.  } 
\end{figure}

We are aware that historical perspective might sound inappropriate in this type of research but we will try to justify the need for this section. The idea of using the cell phone with camera and (large) screens started to appear around the early 2000s. One of the novel ideas back then was the \textit{screen registration} \cite{registered_screen_proof} which is the most similar to our idea however upon studying it in detail we understand that it is completely backwards compared to our approach. Namely, in the screen registration approach either the cell phone sends the data about recognized markers to the server or it even sends the whole captured image from the cell phone to the server. Server then calculates its own position in the camera framework of the client. After development of our technique we were actually quite surprised to discover unusual approach used in the screen registration technique which obviously has several drawbacks -- the most obvious one being unnecessary communication with the server. To understand this we believe that the origin of the approach was in the technical capabilities of the cell phones when it was developed. This is the historical perspective since the last 10 years in the development of the smartphones were so intense.\\
  The original paper presenting the screen registration technique \cite{registered_screen_proof} was done when the cell phones were not even smartphones (2008). They were rather weak machines with 100 MHz CPUs and with no touchscreens. On the other hand camera resolutions were small and network speed was reasonable. So the obvious idea to do some computation is to grab the (relatively small resolution) image and to send it to the server which could be a powerful machine even back then. The server can then perform more complex operations such as searching for image features (in order to achieve markerless solution) \cite{touch_projector_1,touch_projector_2}. With the advance of the smartphones (2011) screen registration was implemented \cite{touch_projector_1} using the fiducial markers, even though using the lighted edge of the facade is very similar to our idea, but again implemented backwards. The smartphone sending the image to the server which then searches for the fiducial markers (lines) and determines the relative position of the user's phone with respect to the screen. Then user can touch the cell phone where the (big) screen is visible in the video of the phone and perform drawing on the big screen since the server can calculate its coordinates within the phones screen. The idea was for the user to have a feeling as if touching the big screen directly. This was called \textit{Touch Projector}\cite{touch_projector_1,touch_projector_2} and it brings us to another historical point in the development of the smartphones and web technology in general in the last 10 years. Such application of the smartphone to draw on the common big screen would today be implemented trivially. Since the smartphone has to establish connection with the server (screen) anyways many users can draw on a common screen and see the produced image on their own phones instantly without any need to hold their phone in the air aiming the camera to the screen. This is exactly how any tool for collaborative work operates today and maybe the best examples are the massive multiplayer games which became overwhelming in the last 5 years - Minecraft \cite{website_MineCraft}, and AgarIO \cite{website_agarIO} to mention the two. In that way users can work on the same \textit{virtual} screen together and of course server can show it on a big screen. The result is that many applications that envisioned interaction between the screens and smartphones are simply obsolete today due to advances in cloud technologies etc. Notice that our implementation in particular is based on recalculating the position of the center point of the user screen (0.5, 0.5) and sending it to the server but our approach can be trivially turned into a Touch Projector interface. Nevertheless, there is one type of interaction that still makes sense regarding the screen smartphone interaction and that is turning a smartphone into a direct pointing device for the screen. We believe that this is also the reason that in the arcade shops - the very few left - the arcades that one can find there are pinball and the \textit{Operation Wolf} \cite{website_opwolf} type of games where the players shoot at the screen using the attached weapons (crossbows, machine guns, etc.). The reason for this is that this kind of experience is still not available without the special equipment no matter how advanced the smartphone technology today is. Our implementation offers exactly this type of interaction and is perfectly suited for many players on the same screen. In order to recreate such experience we made a simple device for holding a smartphone on a toy gun shown in Fig. \ref{toy_gun}.
  
The cell phone attached to a toy gun in this way was recreating the original DuckHunt playing experience even better than the original. Unlike the original DuckHunt that could only check if the duck was hit or not with this toy gun user can aim and shoot anywhere at the screen.\\

\begin{figure}[ht]
\centering
\includegraphics[clip=true,width=0.95\textwidth]{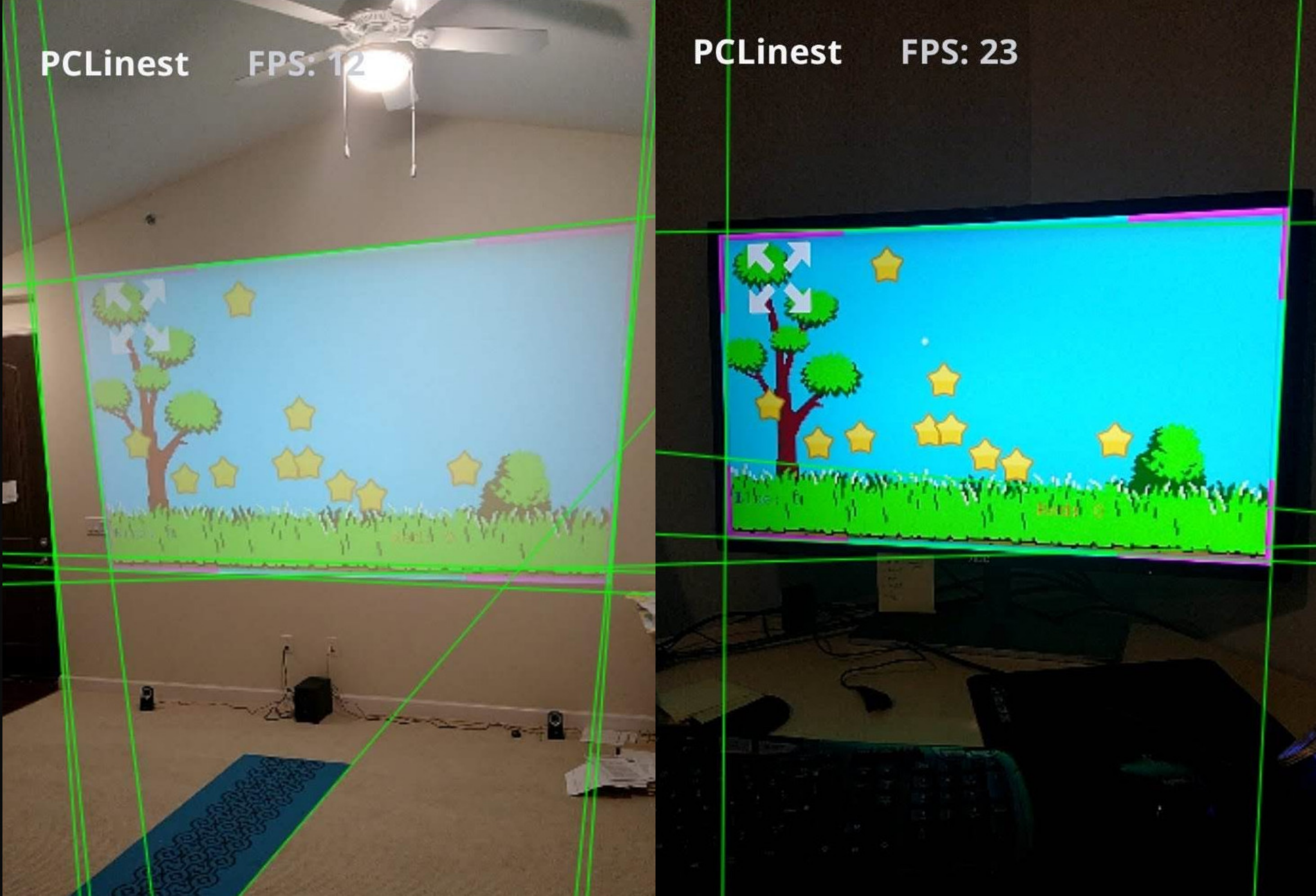}
\caption{\label{fig7} Testing of finding the PCLines running the GammaCV library. Notice that some of the lines are falsely detected features either from the game content or from the objects in the surroundings of the screen. Even though we did not implement it, our idea in finding the proper lines would be looking for lines that resemble a rectangle (i.e. making right angles between them) and then selecting the four which give a rectangle of the proper height--width ratio which was reported from the server.} 
\end{figure}

Since we are referencing advances of technology here we should mention that further step in improving our approach would be using markerless screen detection and for that purpose we have used the GammaCV library \cite{website_GammaCV} which is highly optimized for finding fiducial markers \cite{website_FiducialMarker} such as PCLines \cite{PC_Lines} and Canny Edges \cite{website_CannyEdge} using the smartphone GPU it is still almost unusable on the relatively high end cellphone. We performed tests on one of the smartphones and the lines detection was done at 11 FPS, which even if accepted as playable, would still drain the cell phone battery rather quickly. The example of lines detection is shown in Fig. \ref{fig7}. With this we could indeed implement a truly markerless solution of our algorithm but at a very high cost in GPU/CPU demand. Therefore we find the multicolor edge as an optimal tradeoff in losing some of the screen area but having acceptably fast algorithm performed on the smartphone's CPU/GPU.\\      

Finally, it is worth mentioning that our approach does not necessarily require interaction with the computer screen. Once the cell phone established the communication with server the interaction can be achieved through the camera and the stationary billboard or even with the building facade which is controlled by the server \cite{touch_projector_1}. One can think of several other original ways of interacting with physical objects.

\section{Conclusions}

\label{concl}

We have demonstrated that a simple idea of recalculating screen coordinates from the camera streaming video can be turned into a powerful interaction technique which provides seamless interaction between the users with smartphone and a (large) screen. Moreover, the technique has such a low demand regarding the network throughput that it can be used by literally hundreds of users on the same screen providing a completely novel and original ways of using the smartphone.\\ 
We believe that this technique will catch up and become very popular with the raise of the HTML5 and improvement of browsers. We hope it will spark the creation of completely novel types of massive multiplayer games on common screens.\\
Finally, we hope that the readers have enjoyed playing the game \cite{mobiletvgames} that accompanies this paper.

\section{Acknowledgements}
The Author thanks Ana Babi\'{c} and Timothy Middelkoop for fruitful discussion.

%
%


\bibliographystyle{unsrt}

\bibliography{duck_hunt_references}   

\end{document}